\documentclass[12pt,a4paper,sort&compress]{article}
\usepackage{jheppub}
\usepackage{amsmath}
\usepackage{amssymb}
\usepackage{amsfonts}
\usepackage{amsthm}
\usepackage{mathtools}
\usepackage{stmaryrd}
\usepackage{shuffle}
\usepackage{enumitem}
\usepackage{color}   
\usepackage{hyperref}
\usepackage{adjustbox}
\usepackage{graphicx}
\usepackage[dvipsnames]{xcolor}
\colorlet{LightAquamarin}{Aquamarine!30}
\colorlet{LightRubineRed}{RubineRed!30}
\definecolor{UniBlau}{RGB}{7,82,154}
\usepackage{framed}
\usepackage{tikz}
\usepackage{bbold}

\newcommand{\eps}{\epsilon}
\newcommand{\ord}{\begin{cal}O\end{cal}}

\newcommand{\rd}{\mathrm{d}}

\def\beq{\begin{equation}}
\def\eeq{\end{equation}}
\def\bsp#1\esp{\begin{split}#1\end{split}}

\newenvironment{sloppyequation}[0]{\sloppy\begin{flushleft}\hspace*{0.75cm}\(}{\)\end{flushleft}\fussy}

\newcommand{\beqsloppy}{\begin{sloppyequation}}
\newcommand{\eeqsloppy}{\end{sloppyequation}}

\newcommand{\cA}{\begin{cal}A\end{cal}}

\newcommand{\cD}{\begin{cal}D\end{cal}}

\newcommand{\cN}{\begin{cal}N\end{cal}}

\usepackage{hyperref}
\theoremstyle{definition}

\newtheorem{thm}{Theorem}

\newcommand{\bK}{{\textbf K}}
\newcommand{\bR}{{\textbf R}}
\newcommand{\bF}{{\textbf F}}

\newcommand{\bW}{{\overline W}}

\newcommand{\bU}{{\textbf U}}

\newcommand{\bRbar}{\overline{\textbf R}}

\newcommand{\tR}{{{R_2}}}
\newcommand{\cAbar}{\overline{\mathcal A}}
\newcommand{\cNbar}{\overline{\mathcal N}}
\newcommand{\Phibar}{\overline{\Phi}}
\newcommand{\Phibarzero}{\Phibar_R\!\!\!{}^0}

\DeclareMathOperator{\id}{id}

\newcommand{\MS}{\overline{\textrm{MS}}}

\allowdisplaybreaks

\title{Rational terms of UV origin to all loop orders}

\author[a]{Claude Duhr}
\emailAdd{cduhr@uni-bonn.de}

\author[a]{Paarth Thakkar}
\emailAdd{pthakkar@uni-bonn.de}

\affiliation[a]{Bethe Center for Theoretical Physics, Universit\"at Bonn, D-53115, Germany}

\abstract{Numerical approaches to computations typically reconstruct the numerators of Feynman diagrams in four dimensions. In doing so, certain rational terms arising from the $(D-4)$-dimensional part of the numerator multiplying ultraviolet (UV) poles in dimensional regularisation are not captured and need to be obtained by other means. At one-loop these rational terms of UV origin can be computed from a set of process-independent Feynman rules. Recently, it was shown that this approach can be extended to two loops. In this paper, we show that to all loop orders it is possible to compute rational terms of UV origin through process-independent vertices that are polynomial in masses and momenta.}

\begin{document}

\preprint{BONN-TH-2023-12}

\maketitle


\section{Introduction}
\label{sec:intro}

Performing precise predictions for high-energy collider experiments, like the Large Hadron Collider (LHC) at CERN, typically requires one to evaluate Feynman diagrams involving loops of virtual particles. In this context, the computation of one-loop corrections is often considered a solved problem, because there are several computer codes that allow one to evaluate one-loop corrections in an automated way, e.g., refs.~\cite{Berger:2008sj,Bevilacqua:2011xh,Cullen:2011ac,Hirschi:2011pa,Actis:2016mpe,Biedermann:2017yoi,Buccioni:2019sur}. Many of these codes rely, in one way or another, on unitarity-based approaches~\cite{Bern:1994cg,Bern:1994zx} and the OPP method to reconstruct the numerators of one-loop Feynman diagrams~\cite{Ossola:2006us}. Many of these algorithms only reconstruct numerators of Feynman diagrams with all momenta and Dirac matrices in four space-time dimensions. Feynman integrals typically diverge, and the most common regularisation is dimensional regularisation, where the integrals are evaluated in $D=4-2\eps$. The difference between the results obtained with the $D$- and four-dimensional numerators are purely rational terms at one-loop. There are two classes of rational terms. The so-called $R_1$ terms can be obtained during the procedure of reconstructing the numerator. The $R_2$ terms instead arise from poles in $\epsilon$ multiplying the $(D-4)$-dimensional part of the numerator. An independent method is thus needed to reconstruct the rational $R_2$ terms. 
At one loop all $R_2$ terms stem from poles of ultraviolet (UV) origin~\cite{Bredenstein:2008zb}, and they can be computed using a set of process-independent tree-level Feynman rules that are polynomial in momenta and masses~\cite{Ossola:2008xq}. This allows one to implement these Feynman rules into automated computer codes (e.g., using the UFO format~\cite{Darme:2023jdn}) and to compute the $R_2$ terms from tree-level Feynman diagrams. At one loop, these Feynman rules have been evaluated for processes within the Standard Model~\cite{Draggiotis:2009yb,Garzelli:2009is,Pittau:2011qp} and beyond~\cite{Degrande:2014vpa,Degrande:2020evl}.

In order to reach the precision required by the experiments, one-loop computations are not enough, and also higher-loop Feynman diagrams need to be evaluated. There was a lot progress in recent years in understanding the structure of multi-loop Feynman integrals (see, e.g., refs.~\cite{Bourjaily:2022bwx,Caola:2022ayt} for recent reviews), and numerical unitarity methods at two loops are also slowly reaching maturity (see, e.g., refs.~\cite{Ita:2015tya,Abreu:2020xvt}). So far, however, these advances are restricted to computations with few external legs. For the future, automated approaches to two- and higher-loop computations, like at one-loop, are highly needed and desirably, and first steps in this direction have already been taken~\cite{Pozzorini:2022msb,Pozzorini:2022ohr,Zoller:2022ewt}. These approaches also reconstruct the numerators of Feynman diagrams in four dimensions, so again one needs to recover the $R_2$ terms by other means. It was recently shown that also at two loops one can define a set of process-independent tree-level Feynman rules that are polynomial in masses and momenta so that the rational $R_2$ terms of UV origin can be obtained by computing all tree-level and one-loop diagrams with these new vertices inserted~\cite{Lang:2020nnl,Lang:2021hnw,Pozzorini:2020hkx}. 

The purpose of this paper is to address the Quantum Field Theory question in how far it is possible to obtain tree-level Feynman rules to compute rational $R_2$ terms of UV origin even at higher loops. We will show that it is indeed possible to define a set of vertices that are polynomial in momenta and masses, such that the Feynman diagrams with these vertices inserted compute precisely the UV counterterms and the $R_2$ terms needed to obtain the value of the Feynman diagram in the (modified) minimal subtraction ($\MS$) scheme. Our result shows that the definition of the Feynman rules for $R_2$ terms of UV origin has a field theoretic definition to all orders in perturbation theory.

This paper is structured as follows. In section~\ref{sec:dimreg} we introduce our notations and conventions, and in section~\ref{sec:renormalisation} we review UV renormalisation in the $\MS$-scheme. In section~\ref{sec:hopf_approach} we present our definition of rational $R_2$ terms, and we show that can always be interpreted in terms of Feynman diagrams to all orders in perturbation theory. Finally, in section~\ref{sec:conclusion} we draw our conclusions. We include an appendix where we collect the mathematical proofs of our main results.


\section{Multi-loop amplitudes in dimensional regularisation}
\label{sec:dimreg}

Our goal is to study $L$-loop Feynman integrals in Quantum Field Theory (QFT) attached to a Feynman graph $\Gamma$ with $n$ external legs. More precisely, we are interested in Feynman diagrams $\Gamma$ that contribute to  amputated off-shell Green's functions,\footnote{We will often refer to $\cAbar_{\Gamma}$ as `amplitude', despite the fact that it is the analytic expression of a single Feynamn diagram $\Gamma$ contributing to an amputated off-shell Green's function.}
\beq\label{eq:int_Def}
\cAbar_{\Gamma}(p_1,\ldots,p_n) = \int\left(\prod_{i=1}^L\frac{\rd^D\bar{q}_i}{(2\pi)^D}\right)\,\frac{\cNbar_{\!\Gamma}(p_i,\bar{q}_j)}{\prod_{e\in E_\Gamma}\overline{D}_e}\,, \qquad \sum_{i=1}^np_i=0\,,
\eeq
where $E_{\Gamma}$ denotes the set of edges of the graph $\Gamma$ and $\overline{D}_e$ is the denominator of the propagator attached to the edge $e$:
\beq\label{eq:pro_Def}
\overline{D}_e = \left(\sum_{i=1}^n\alpha_{e,i}\,p_i+\sum_{j=1}^L\beta_{e,j}\,\overline{q}_j\right)^{\!\!2} - m_e^2\,,\qquad \alpha_{e,i}\,,\beta_{e,j}\in\{-1,0,+1\}\,.
\eeq
Note that, if we allow graphs with two-point vertices, this definition encompasses propagators raised to higher powers.
The numerator $\cNbar_{\!\Gamma}(p_i,\bar{q}_j)$ is a polynomial in the external and internal momenta $p_i$ and $\bar{q}_i$ and the masses $m_e$ of the propagators. In the case of external particles with spins greater than zero (typically, Dirac fermions and/or vector bosons), the numerator is actually a tensor in the corresponding spin or Lorentz indices. We will suppress this tensorial structure in the following.

The integral in eq.~\eqref{eq:int_Def} may diverge. In order to regulate the divergences we work in dimensional regularisation in $D=4-2\epsilon$ dimensions. More specifically, we will work in the `t\,Hooft-Veltman (HV) scheme, where all external momenta and polarisations are chosen to lie in four dimensions, and all internal momenta $\bar{q}_j$ lie in $D$ dimensions. We denote $D$-dimensional quantities with a bar. Their projections to four dimensions are represented without the bar, and the $(D-4)$-dimensional components are denoted with a tilde. For example, the $D$-dimensional loop momenta are $\bar{q}_j^{\bar{\mu}} = q_j^{\mu}+\tilde{q}_j^{\tilde{\mu}}$, where $q_j^{\mu}$ and $\tilde{q}_j^{\tilde{\mu}}$ are their four- and $(D-4$)-dimensional components. $D$-dimensional Lorentz indices are contracted with the $D$-dimensional metric $\bar{g}_{\bar{\mu}\bar{\nu}}$:
\beq
\bar{a}_{\bar{\mu}}\bar{b}^{\bar{\mu}} = \bar{g}_{\bar{\mu}\bar{\nu}}\,\bar{a}^{\bar{\nu}}\,\bar{b}^{\bar{\mu}}={a}_{{\mu}}{b}^{{\mu}}+\tilde{a}_{\tilde{\mu}}\tilde{b}^{\tilde{\mu}}\,,
\eeq
where the metric tensors in the four- and $(D-4)$-dimensional spaces are defined by
\beq
{a}_{{\mu}}{b}^{{\mu}} = {g}_{{\mu}{\nu}}\,{a}^{{\nu}}\,{b}^{{\mu}} \textrm{~~~and~~~}\tilde{a}_{\tilde{\mu}}\tilde{b}^{\tilde{\mu}}=\tilde{g}_{\tilde{\mu}\tilde{\nu}}\,\tilde{a}^{\tilde{\nu}}\,\tilde{b}^{\tilde{\mu}}\,.
\eeq
Note that, if the Feynman graph $\Gamma$ contains fermion lines, the numerator will typically involve $D$-dimensional Dirac matrices $\bar{\gamma}^{\bar{\mu}}$, and $D$-dimensional identities between Dirac matrices will have to be used.\footnote{The integrand may also involve the Dirac matrix $\gamma^5$ and/or the Levi-Civita tensor $\epsilon^{\mu\nu\rho\sigma}$. Both are purely four-dimensional objects, and care needs to be taken how to implement them into dimensional regularisation. We do not expect that any of our conclusions rely on this issue in any essential way, but we exclude them from our discussion, since their treatment requires special care, see, e.g.,~ref.~\cite{Shao:2011tg}.}

In applications, especially when automated computer tools to generate Feynman diagrams are involved, it is often easier to generate the numerators in four rather than $D$ dimensions. This then leads to the integral
\beq\label{eq:int_Def4}
\cA_{\Gamma}(p_1,\ldots,p_n) = \int\left(\prod_{i=1}^L\frac{\rd^D\bar{q}_i}{(2\pi)^D}\right)\,\frac{\cN_{\!\Gamma}(p_i,{q}_j)}{\prod_{e\in E_\Gamma}\overline{D}_e}\,.
\eeq
Note that the integration measure, as well as the denominators of the propagators, are kept $D$-dimensional. The numerator, instead, is purely four-dimensional, and can be obtained from the numerator $\cNbar_{\!\Gamma}(p_i,\bar{q}_j)$ in eq.~\eqref{eq:int_Def} by simple projection to four dimensions:
\beq
\cN_{\!\Gamma}(p_i,{q}_j) = \cNbar_{\!\Gamma}(p_i,\bar{q}_j)_{\big|D\to4,\,\bar{g}_{\bar{\mu}\bar{\nu}}\to {g}_{{\mu}{\nu}},\,\bar{\gamma}^{\bar\mu}\to{\gamma}^{\mu}}\,.
\eeq
If the integral in eq.~\eqref{eq:int_Def} is convergent, then it is equal to the integral computed with the numerator in four dimensions, up to terms that vanish in four dimensions, i.e., $\cAbar_{\Gamma} = \cA_{\Gamma} + \ord(\epsilon)$ for convergent integrals. However, if the integral in eq.~\eqref{eq:int_Def} is divergent in $D=4$ dimensions, the difference $\cAbar_{\Gamma} - \cA_{\Gamma}$ may involve finite contributions, and possibly even poles. Hence, a method is required to evaluate this difference separately. 

In ref.~\cite{Ossola:2008xq} it was shown that for one-loop graphs, the difference $\cAbar_{\Gamma} - \cA_{\Gamma}$ is a purely rational function of the external momenta and masses. More precisely, it is possible to define a novel set of vertices, called $R_2$\emph{-vertices}, which are polynomial in the momenta and masses, and the finite contributions to the difference $\delta R_{\Gamma}=\cAbar_{\Gamma} - \cA_{\Gamma}$ are obtained by considering all possible tree-level Feynman graphs with a single $R_2$ vertex inserted. Recently, it was shown that a similar approach can be used at two loops~\cite{Pozzorini:2020hkx}: it is possible to define a set of two-loop $R_2$ vertices, which are polynomial in the momenta and masses, and the difference between the (renormalised) amplitudes computed with $D$- and four-dimensional numerators is obtained by considering all one-loop and tree-level  Feynman graphs with one- and two-loop $R_2$ vertices inserted:
\beq\bsp\label{eq:two-loop-R2-master}
\bR\cAbar_{\Gamma} &\,= \cA_{\Gamma} +\delta R_{\Gamma}+\delta Z_{\Gamma} +\sum_{\gamma}(\delta R_{\gamma}+\delta Z_{\gamma})\ast \cA_{\Gamma/\gamma}+\ord(\eps)\,,
\esp\eeq
where the sum runs over all one-particle irreducible (1PI) subgraphs of $\Gamma$.
$\bR\cAbar_{\Gamma} $ is the renormalised diagram in the $\MS$ scheme, and $\delta Z_{\gamma}$ and $\delta Z_{\Gamma}$ are the corresponding one- and two-loop UV counterterms. More precisely, at two loops we have:\footnote{In ref.~\cite{Pozzorini:2020hkx} the UV counterterm is written as $\delta Z_{\Gamma}+\delta \widetilde{Z}_{\Gamma}$, where $\delta Z_{\Gamma}$ captures the four-dimensional contributions, and $\delta \widetilde{Z}_{\Gamma}$ depends on the $(D-4)$-dimensional components of the loop momentum. In our conventions, the UV counterterm is the sum of these two contributions.} The quantity $\delta R_{\gamma}$ is the one-loop $R_2$ vertex for the one-loop subgraph $\gamma$, and $\delta R_{\Gamma}$ is the two-loop $R_2$ vertex, which is a polynomial in internal masses and external momenta. The $R_2$ vertices are non-zero only for superficially UV divergent graphs (see the next section), and can be computed from eq.~\eqref{eq:two-loop-R2-master}
\beq\bsp
\delta R_{\Gamma} &\,=\bR\cAbar_{\Gamma}- \cA_{\Gamma} -\delta Z_{\Gamma} -\sum_{\gamma}(\delta R_{\gamma}+\delta Z_{\gamma})\ast \cA_{\Gamma/\gamma}+\ord(\eps)\,.
\esp\eeq

Let us conclude this review by commenting on infrared divergencies. If the QFT under consideration involves massless particles, then the loop amplitudes will typically have poles of both ultraviolet (UV) and infrared (IR) origin. At one loop, it is known that IR singularities can generate finite contributions to the difference $\cAbar_{\Gamma} - \cA_{\Gamma}$ in intermediate steps of the computations, but these contributions cancel in the final result in the HV scheme~\cite{Bredenstein:2008zb}. At two loops, the effect of IR singularities on two-loop rational $R_2$ terms is still an open question, and the results of ref.~\cite{Pozzorini:2020hkx} are strictly speaking only applicable to rational $R_2$ of UV origin.\footnote{There are indications that rational terms of IR origin cancel at two loops at least in QED~\cite{Zhang:2022rft}.}

Beyond two loops, it is currently unknown if the finite contributions to the difference $\cAbar_{\Gamma} - \cA_{\Gamma}$ can be described by a set of $R_2$ Feynman rules that are polynomial in momenta and masses. In the remainder of this paper we show that, at least for those contributions that arise from UV poles, this is indeed the case. Before we discuss our generalisation of the results of ref.~\cite{Pozzorini:2020hkx} to arbitrary orders, we need to have a closer look at the structure of UV singularities in QFT.


\section{UV renormalisation and the $\MS$-scheme}
\label{sec:renormalisation}

\subsection{Renormalised amplitudes and the forest formula}
\label{sec:BPHZ}
In this section we review the renormalisation of QFT amplitudes in the $\MS$-scheme. Since we are only interested in UV singularities, we restrict the discussion to theories without IR singularities, though the conclusion of the structure of UV singularities does not depend on this restriction on any essential way. We also restrict the discussion to the amplitude $\cAbar_{\Gamma}$, but all the formulas equally apply to $\cA_{\Gamma}$. 

We start by reviewing \emph{Weinberg's theorem}, which gives a sufficient condition to decide if a Feynman integral (in Euclidean space, i.e., after Wick rotation) is convergent~\cite{Weinberg:1959nj}. We define in the usual way the \emph{superficial degree of divergence} of $\Gamma$ as the naive scaling of the integrand as all loop momenta go to infinity simultaneously,
\beq
\omega(\Gamma) = LD - 2\,N_{\textrm{prop}} + \deg \cNbar\,,
\eeq
where $N_{\textrm{prop}}$ denotes the number of propagators that involve at least one loop momentum and $\deg \cNbar$ is the degree of the numerator $\cNbar$ seen as a polynomial in the loop momenta $\bar{q}_j$. Weinberg's theorem then states that, if $\omega(\Gamma)<0$ and $\Gamma$ does not contain any divergent subgraph, then the integral $\cAbar_{\Gamma}$ converges (in Euclidean space). We stress that Weinberg's theorem only expresses a sufficient condition for convergence, but it is by no means necessary.

UV singularities are recursively removed by the renormalisation procedure. The subtraction is encoded in the Bogoliubov-Parasiuk-Hepp-Zimmermann (BPHZ) forest formula~\cite{Bogoliubov:1957gp,Hepp:1966eg,Zimmermann:1969jj}. The renormalised amplitude is given by
\beq\label{eq:R_def}
\bR\cAbar_{\Gamma} = (1-\bK)\Big[\cAbar_{\Gamma}-\sum_{\gamma\in W(\Gamma)}\bRbar\,\cAbar_{\gamma}\ast\cAbar_{\Gamma/\gamma}\Big]\,.
\eeq
The operator $\bK$ extracts the poles in $\epsilon$, and its precise form defines the renormalisation scheme. For the Minimal Subtraction (MS) and Modified Minimal Subtraction ($\MS$) schemes its action on a Laurent series $f(\eps) = \sum_{k\ge k_0} f_k\,\eps^k$ is given by
\beq\bsp\label{eq:K_def}
\bK_{\textrm{MS}}f(\eps) &\,= \sum_{k=k_0}^{-1}f_k\,\eps^k\,,\\
 \bK_{{\MS}}f(\eps) &\,= e^{L\gamma_E\eps}(4\pi)^{L\eps}\,\bK_{\textrm{MS}}\Big[e^{-L\gamma_E\eps}(4\pi)^{-L\eps}\,f(\eps)\Big]\,,
\esp\eeq
with $L$ the number of loops of the graph it acts on, and $\gamma_E=-\Gamma'(1)$ is the Euler-Mascheroni constant. 
Finally, $\bRbar$ defines the UV counterterm (in the scheme defined by $\bK$) of the subgraph $\gamma$:\footnote{Note that the bar on top of $\bRbar$ does not refer to $D$ vs. four-dimensional components.}
\beq\label{eq:Rbar_def}
\bRbar\,\cAbar_{\Gamma} = \bK\Big[\cAbar_{\Gamma}-\sum_{\gamma\in W(\Gamma)}\bRbar\,\cAbar_{\gamma}\ast\cAbar_{\Gamma/\gamma}\Big]\,.
\eeq
where $W(\Gamma)$ is the set of all products of 1PI subgraphs of $\Gamma$ that are different from the empty graph and from $\Gamma$ itself. $\Gamma/\gamma$ denotes the Feynman graph obtained by shrinking the subgraph $\gamma$ to a point, and the insertion operator is defined by
\beq
f_{\gamma}\ast \cAbar_{\Gamma/\gamma} = \int\left(\prod_{i=1}^{L'}\frac{\rd^D\bar{q}_i}{(2\pi)^D}\right)\,f_{\gamma}(p_i,\bar{q}_i)\,\frac{\cNbar_{\!\Gamma/\gamma}(p_i,\bar{q}_j)}{\prod_{e\in E_{\Gamma/\gamma}}\overline{D}_e}\,,
\eeq
with $L'$ the number of loops of $\Gamma/\gamma$. 
Since $\bK(1-\bK)=0$, the renormalised amplitude is always finite. Note that the renormalised amplitude in eq.~\eqref{eq:R_def} can also be written
\beq\label{eq:R_to_Rbar}
\bR\cAbar_{\Gamma} = \sum_{\gamma\in \overline{W}(\Gamma)} \delta \overline{Z}_{\gamma} \ast\cAbar_{\Gamma/\gamma}\,,
\eeq
where we introduced the UV counterterm of a graph $\gamma$:
 \beq
 \delta \overline{Z}_{\gamma} = -\bRbar\,\cAbar_{\gamma}\,.
 \eeq
The sum runs over \emph{all} products of 1PI subgraphs, including the empty graph $\emptyset$ and $\Gamma$ itself, and we define $\bW(\Gamma) = W(\Gamma)\cup\{\emptyset,\Gamma\}$, and $\Gamma/\Gamma =\emptyset$ and $\Gamma/\emptyset=\Gamma$.

 In the MS and $\MS$-schemes, the maps $\bR$ and $\bRbar$ have the following properties:
\begin{enumerate}
\item $\bR$ and $\bRbar$ are multiplicative:
\beq\bsp
\bR\big[\cAbar_{\Gamma_1}\cdot\cAbar_{\Gamma_2}\big]&\, = \bR\cAbar_{\Gamma_1}\cdot\bR\cAbar_{\Gamma_2}\,,\\ 
\bRbar\big[\cAbar_{\Gamma_1}\cdot\cAbar_{\Gamma_2}\big] &\,= \bRbar\,\cAbar_{\Gamma_1}\cdot\bRbar\,\cAbar_{\Gamma_2}\,.
\esp\eeq
\item If $\omega(\Gamma)<0$, then $\bRbar\,\cAbar_{\Gamma}=0$.
\item $\bR$ and $\bRbar$ commute with derivatives with respect to external momenta or propagator masses.
\item $\bRbar\,\cAbar_{\Gamma}$ is a polynomial in external momenta and propagator masses.
\end{enumerate}
The proofs of these statements can for example be found in ref.~\cite{Caswell:1981ek}. 
 
 Let us discuss how the aforementioned properties of the $\bR$ and $\bRbar$ maps lead to a Feynman diagrammatic interpretation. 
 Due to Properties 1 and 2, we only need to compute UV counterterms for superficially divergent 1PI graphs. In a renormalisable theory, there are typically only a finite number of superficially divergent 1PI graphs at a given loop order. Property 4 implies that the UV counterterms of these superficially divergent 1PI graphs are polynomials, and can thus be interpreted as tree-level Feynman rules. The renormalised amplitude $\bR\cAbar_{\Gamma}$ is then obtained by adding to $\cAbar_{\Gamma}$ the Feynman graphs where we have replaced in all possible ways superficially divergent 1PI subgraphs $\gamma$ by their UV counterterms $\delta \overline{Z}_{\gamma}$. We see that properties 1, 2 \& 4 are sufficient to achieve a diagrammatic interpretation of the counterterms (Property 3 is needed in order to prove Property 4). 
 
We conclude with some comments. First, we work in the HV scheme, where all external momenta are taken in four dimensions. The UV counterterms, however, should be evaluated with $D$-dimensional external momenta. Indeed, if $\gamma$ appears as a subgraph of a higher-loop graph $\Gamma$, then the external legs of $\gamma$ may contain loop momenta, which are $D$-dimensional. Second, we can apply the same reasoning to the amplitude $\cA_{\Gamma}$, and its renormalisation requires UV counterterms $\delta Z_{\gamma}$, which may be different from the UV counterterms $\delta \overline{Z}_{\gamma}$ obtained from $\cAbar_{\Gamma}$. The counterterm vertices $\delta Z_{\gamma}$ will be used to construct numerators for the counterterm graphs, but they have a dependence on the $(D-4)$-dimensional components of the loop momenta, stemming from the $D$-dimensional momenta in the propagators in $\cA_{\gamma}$. We refer to ref.~\cite{Pozzorini:2020hkx} for a detailed discussion at two-loop order.

\subsection{Hopf algebraic approach to renormalisation}
\label{sec:Hopf}
UV renormalisation is mathematically captured by a Hopf algebra structure on graphs~\cite{Connes:1999yr} (see also ref.~\cite{Manchon:2001bf} for a review). In the remainder of this section we review the Hopf algebraic approach to renormalisation, as it provides a language that we can use to study $R_2$-terms in subsequent sections. We start by giving a concise review of the Hopf algebra of graphs, before we discuss the application to UV renormalisation. Note that we only review the basics necessary to understand UV renormalisation, and how to extend these concepts to $R_2$ terms. For a more complete mathematicla review, see,~e.g.,~ref.~\cite{Manchon:2001bf}.

\paragraph{The Hopf algebra of graphs.} In the following a \emph{graph} denotes a triplet $(E_{\Gamma},V_{\Gamma},\phi)$, where $E_{\Gamma}$ is a set of \emph{edges}, $V_{\Gamma}$ a set of vertices, and $\phi$ is a map that assigns to each edge its endpoints in $V_{\Gamma}$. If $v\in V_{\Gamma}$ is an endpoint of $e\in E_{\Gamma}$, we say that $v$ is incident to $e$. The valency of a vertex is the number of distinct edges to which it is incident. Note that we allow graphs with external edges, i.e., edges with valency 1. 

Let $H$ denote the $\mathbb{Q}$-algebra generated by all graphs $\Gamma$ (including the empty graph), where the multiplication is simply the disjoint union of graphs. $H$ can be equipped with a coproduct $\Delta:H\to H\otimes H$ defined by
\beq
\Delta(\Gamma) := 1\otimes \Gamma + \Gamma\otimes 1 + \sum_{\gamma\in W(\Gamma)} \gamma\otimes (\Gamma/\gamma)\,,
\eeq
where $W(\Gamma)$ and $\Gamma/\gamma$ have been defined below eq.~\eqref{eq:Rbar_def}. $H$ equipped with this coproduct is in fact a Hopf algebra. The counit is simply the augmentation map, i.e., the projection $\varepsilon$ onto the empty graph. The antipode is determined uniquely by the condition
\beq
m(S \otimes \id)\Delta = \varepsilon\,,
\eeq
where $m$ denotes the multiplication in $H$.
From this relation, the antipode can  determined recursively:
\beq
S(\Gamma) = -\Gamma - \sum_{\gamma\in W(\Gamma)}S(\gamma)(\Gamma/\gamma)\,,\qquad S(1) = 1\,.
\eeq

Let $A$ be an algebra over some ring $R$ (for us typically $R=\mathbb{C}$. A \emph{character} with values in $A$ is an algebra homomorphism from $H$ to $A$, i.e., a linear map $\phi$ that preserves the multiplication:
\beq
\phi(\Gamma_1\Gamma_2) = \phi(\Gamma_1)\phi(\Gamma_2)\,.
\eeq
If $H$ is a Hopf algebra, the set of all characters forms a group for the convolution product
\beq
\phi_1\star \phi_2 := m_A(\phi_1\otimes\phi_2)\Delta\,,
\eeq
where $m_A$ denotes the multiplication in $A$. The unit for the convolution product is the counit $\varepsilon$, and the inverse for the convolution is obtain by composing with the antipode, $\phi^{\star-1} := \phi S$.
Finally, a \emph{Rota-Baxter map} is a map $\kappa:A\to A$ such that, for all $x,y\in A$, 
\beq
\kappa(x)\kappa(y) = \kappa(\kappa(x)y) + \kappa(x\kappa(y)) -\kappa(xy)\,.
\eeq
One can show that, given a Rota-Baxter map $\kappa$, every character $\varphi$ with values in $A$ admits a \emph{Birkhoff decomposition}, i.e., it can be written in the form
\beq
\varphi = \varphi_-^{\star-1} \star \varphi_+\,.
\eeq
The characters $\varphi_{\pm}$ can be explicitly written down in a recursive manner:
\beq\bsp\label{eq:Birkhoff}
\varphi_- &\,= -\kappa\left[\varphi + m_A(\varphi_-\otimes \varphi)\Delta'\right]\,,\\
\varphi_+ &\,= (1-\kappa)\left[\varphi + m_A(\varphi_-\otimes \varphi)\Delta'\right]\,,
\esp\eeq
where $\Delta' = \Delta-1\otimes\id-\id\otimes1$ is the reduced coproduct.

\paragraph{The Hopf algebra approach to renormalisation.} We now review how the previous concepts can be used to reformulate the process of renormalisation. This approach to renormalisation was pioneered in ref.~\cite{Connes:1999yr}. In the following $\Gamma$ denotes a Feynman graph, i.e., a graph together with the external kinematic data (masses and momenta).

In our case, the algebra $A$ is the algebra of Laurent series in the dimensional regulator $\eps$, i.e., elements of $A$ are Laurent series of the form $\sum_{k\ge k_0} a_k \eps^k$, $k_0\in\mathbb{Z}$. We define two characters on the Hopf algebra of graphs $H$ with values in $A$:
\begin{itemize}
\item The character $\Phibar: H\to A$ assigns to a graph $\Gamma\in H$ the amplitude $\Phibar(\Gamma) = \cAbar_{\Gamma}$ in dimensional regularisation in $D=4-2\eps$ computed with the numerator computed in $D$ dimensions. Here we interpret $\cAbar_{\Gamma}$ as a Laurent series in the dimensional regulator.
\item The character $\Phi: H\to A$ is defined in a similar manner, and assigns to a graph $\Gamma\in H$ the amplitude $\Phi(\Gamma) = \cA_{\Gamma}$ in dimensional regularisation computed with the numerator computed in four dimensions.
\end{itemize}

The maps $\bK_{\textrm{MS}}$ and $\bK_{\overline{\textrm{MS}}}$ define Rota-Baxter maps on $A$. We can then write down a Birkhoff factorisation for $\Phibar$ and $\Phi$. For example, for $\Phi$ we obtain (cf.~eq.~\eqref{eq:Birkhoff})
\beq\bsp\label{eq:Birkhoff_renorm}
{Z} &\,= -\bK\Big[\Phi + m_A(Z\otimes \Phi)\Delta'\Big]\,,\\
\Phi_R &\,= (1-\bK)\Big[\Phi + m_A(Z\otimes \Phi)\Delta'\Big]\,,
\esp\eeq
and we have $\Phi= Z^{\star-1}\star \Phi_R$, or equivalently 
\beq
\Phi_R = Z\star \Phi\,.
\eeq
It will be useful to useful to define
\beq\label{eq:b_def}
b := \Phi + m_A(Z\otimes \Phi)\Delta'\,.
\eeq
Comparing eq.~\eqref{eq:Birkhoff_renorm} to eqs.~\eqref{eq:R_def} and~\eqref{eq:Rbar_def}, we see that we have
\beq
\bR\cA_{\Gamma} = \Phi_R(\Gamma) \textrm{~~~and~~~} \bRbar\cA_{\Gamma} = -Z(\Gamma)\,,
\eeq
and
\beq
(A\star B)(\Gamma) = A(\Gamma)\ast B(\Gamma)\,.
\eeq
The characters $Z$, $\Phi_R$ and $\Phibar_R$ satisfy the same four properties as the $\bR$ and $\bRbar$ maps. In particular, multiplicativity is simply the statement that these maps are characters. Property 3, which expresses the commutativity with derivations, requires some explanation (which will be useful later on).

Let $\Gamma$ be a Feynman graph, $a$ a mass or a component of an external momentum, and $x\in F_{\Gamma}=E_{\Gamma}\cup V_{\Gamma}$ a vertex or an edge of $\Gamma$. Following ref.~\cite{Caswell:1981ek}, we want to introduce a way to discuss derivatives with respect to $a$ at the level of the graphs. We denote by $\delta_a^{(x)}\Gamma$ the Feynman graph such that, when we evaluate it with the Feynman rules $\Phi$ (or $\Phibar$), we insert the derivative with respect to $a$ for the vertex or edge $x$. It is easy to check that $\delta_a^{(x)}\big(\Gamma_1\Gamma_2\big)=\Gamma_2\delta_a^{(x)}\Gamma_1+\Gamma_1\delta_a^{(x)}\Gamma_2$, where we use the convention that $\delta_a^{(x)}\Gamma = 0$ if $x\notin F_{\Gamma}$. The coproduct acts on $\delta_a^{(x)}\Gamma$ via
\beq
\Delta\big(\delta_a^{(x)}\Gamma\big) = \sum_{\substack{\gamma\in \overline{W}(\Gamma) \\ x\in F_{\gamma}}}\delta_a^{(x)}\gamma\otimes (\Gamma/\gamma) + \sum_{\substack{\gamma\in \overline{W}(\Gamma) \\ x\in F_{\Gamma/\gamma}}}\gamma\otimes \delta_a^{(x)}(\Gamma/\gamma)\,.
\eeq
We define~\cite{Caswell:1981ek}
\beq
\delta_a\Gamma := \sum_{x\in F_{\Gamma}}\delta_a^{(x)}\Gamma\,.
\eeq
By definition, we have
\beq
\label{eq:Phi_der}
\partial_a\Phi = \Phi\delta_a \textrm{~~~and~~~} \partial_a\Phibar = \Phibar\delta_a\,.
\eeq
It is easy to see that $\delta_a$ is a derivation,
\beq
\delta_a\big(\Gamma_1\Gamma_2\big)=\Gamma_2\delta_a\Gamma_1+\Gamma_1\delta_a\Gamma_2\,.
\eeq
Moreover, it is also a co-derivation,
\beq
\Delta\delta_a = (\delta_a\otimes\id + \id\otimes \delta_a)\Delta\,.
\eeq
Indeed, since for a given subgraph $\gamma\subseteq \Gamma$, every $x\in F_{\Gamma}$ belongs to either $F_{\gamma}$ or $F_{\Gamma/\gamma}$, we have
\beq\bsp
\Delta\big(\delta_a\Gamma\big) &\,= \sum_{x\in F_{\Gamma}} \Big[\sum_{\substack{\gamma\in \overline{W}(\Gamma) \\ x\in F_{\gamma}}}\delta_a^{(x)}\gamma\otimes (\Gamma/\gamma) + \sum_{\substack{\gamma\in \overline{W}(\Gamma) \\ x\in F_{\Gamma/\gamma}}}\gamma\otimes \delta_a^{(x)}(\Gamma/\gamma)\Big]\\
&\,=\sum_{\gamma\in \overline{W}(\Gamma)} \Big[\sum_{ x\in F_{\gamma}}\delta_a^{(x)}\gamma\otimes (\Gamma/\gamma) + \sum_{x\in F_{\Gamma/\gamma}}\gamma\otimes \delta_a^{(x)}(\Gamma/\gamma)\Big]\\
&\,=\sum_{\gamma\in \overline{W}(\Gamma)} \Big[\delta_a\gamma\otimes (\Gamma/\gamma) + \gamma\otimes \delta_a(\Gamma/\gamma)\Big]\\
&\,=(\delta_a\otimes\id + \id\otimes \delta_a)\Delta(\Gamma)\,.
\esp\eeq
From this it follows that
\beq\bsp\label{eq:Phi_R_der}
\partial_a\Phi &\,= \Phi\delta_a\,,\\
\partial_a\Phibar &\,= \Phibar\delta_a\,,\\
\partial_aZ &\,= Z\delta_a\,.
\esp\eeq
Let us show this for the UV counterterm $Z$. The proofs for the other two characters are similar. We proceed by induction in the number of loops, and the claim is obviously true for a tree-level graph. We have
\beq\bsp
\partial_aZ &\,= -\partial_a\bK \Big[\Phi + m_A(Z\otimes \Phi)\Delta'\Big]\\
&\,= -\bK \Big[\partial_a\Phi + \partial_am_A(Z\otimes \Phi)\Delta'\Big]\\
&\,= -\bK \Big[\Phi\delta_a + m_A(\partial_a\otimes \id+\id\otimes \partial_a)(Z\otimes \Phi)\Delta'\Big]\\
&\,= -\bK \Big[\Phi\delta_a + m_A(\partial_a Z\otimes \Phi+Z\otimes \partial_a\Phi)\Delta'\Big]\\
&\,= -\bK \Big[\Phi\delta_a + m_A(Z\otimes \Phi)(\delta_a\otimes \id+\id\otimes \delta_a)\Delta'\Big]\\
&\,= -\bK \Big[\Phi+ m_A(Z\otimes \Phi)\Delta'\Big]\delta_a \\
&\,=Z\delta_a\,.
\esp\eeq


\section{Feynman rules for rational $R_2$ terms of UV origin}
\label{sec:hopf_approach}

In this section we show that to all orders in perturbation theory it is possible to define Feynman rules that allow one to compute rational terms of UV origin, and we present an explicit formula to compute these Feynman rules, similar to the formula for the UV counterterms in eqs.~\eqref{eq:Rbar_def} and~\eqref{eq:Birkhoff_renorm}. More precisely, we will obtain Feynman rules that combine the UV and $R_2$ Feynman rules in the $\overline{\textrm{MS}}$-scheme into a single set of vertices. 

We start by recursively defining a linear map $U:H\to A$ by $U(1)=1$ and
\beq\label{eq:U_def}
U := \bF\Big[\Phibar_R - \Phi - m_A(U\otimes \Phi)\Delta'\Big]\,,
\eeq
where  the operator $\bF$ is similar to $\bK$, but it extracts in addition the coefficient of $\eps^0$:
\beq
\bF\big[f(\eps)\big] = \eps\,\bK\big[\eps^{-1}\,f(\eps)\big]\,.
\eeq
In the MS and $\MS$-schemes, this operator is
\beq\bsp
\bF_{\textrm{MS}}\big[f(\eps)\big] &\,= \sum_{k=k_0}^0f_k\,\eps^k \,,\\
\bF_{{\MS}}\big[f(\eps)\big] &\,= (4\pi)^{L\eps}e^{L\gamma_E\eps}\,\bF_{\textrm{MS}}\big[(4\pi)^{-L\eps}e^{-L\gamma_E\eps}f(\eps)\big]\,.
\esp\eeq
Just like $\bK$, $\bF$ is a Rota-Baxter map.
Note that $U$ depends on the choice of the renormalisation scheme, through the explicit appearance of the renormalised amplitudes. 

Our goal is to show that $U$ can be used to define a set of Feynman rules for UV counterterms and rational $R_2$ terms in the MS and $\overline{\textrm{MS}}$ scheme. More precisely, we show that $U$ satisfies exactly the same properties as the character $Z$ that defines the UV counterterms (see eq.~\eqref{eq:Birkhoff_renorm}). The first step is to prove the following result, which is the central result of our paper:

\begin{thm}\label{thm:main}\emph{
The linear map $U$ has the following properties.
\begin{enumerate}
\item $U$ is a character.
\item If $\omega(\Gamma)<0$, then $U(\Gamma)=0$.
\item $U$ commutes with derivatives with respect to external momenta or propagator masses: $\partial_aU = U\delta_a$.
\item $U({\Gamma})$ is a polynomial in external momenta and propagator masses.
\end{enumerate}
}
\end{thm}
The proof of this theorem can be found in Appendix~\ref{sec:proof}. Here we only discuss its consequences, and how one can see that we obtain indeed the correct definition of rational $R_2$ terms at all loop orders. 

We start by noting that one can rearrange terms and cast eq.~\eqref{eq:U_def} in the form
\beq\label{eq:Phibar0}
\Phibarzero := \bF\Phibar_R =  \bF\big[U\star \Phi\big]\,.
\eeq
Note that, since $\Phibar_R(\Gamma)$ is always finite, $\Phibarzero$ is also a character.
The map $\Phibarzero$ evaluates to $\Phibarzero(\Gamma) = \bR\,\cAbar_{\Gamma}+\ord(\eps)$, and so we get (with $\bU\cAbar_{\Gamma} := U(\Gamma)$)
\beq\label{eq:master1}
\bR\,\cAbar_{\Gamma} = \sum_{\gamma\in\overline{W}(\Gamma)}\bU\cAbar_{\Gamma}\ast \cA_{\Gamma/\gamma} + \ord(\eps)\,.
\eeq
The previous equation expresses the renormalised amplitude with the numerator computed in $D$ dimensions in terms of amplitudes with numerators computed in four dimensions, with some of the subgraphs replaced by their image under $\bU$. Note that eqs.~\eqref{eq:Phibar0} and~\eqref{eq:master1} are similar in structure to eq.~\eqref{eq:Birkhoff_renorm} and~\eqref{eq:R_to_Rbar} that relate the renormalised amplitudes to the bare amplitudes, but with the numerator computed in the same dimension. Our goal is to interpret $U$ as a set of Feynman rules. Since our theorem shows that the character $U$ satisfies the same properties as the character $Z$ defining the UV counterterms, $\bU\cAbar_{\Gamma} = U(\Gamma)$
 has a direct interpretation in terms of Feynman diagrams, just like for UV renormalisation: The renormalised Feynman diagram $\bR\cAbar_{\Gamma}$ is obtained by adding to $\cA_{\Gamma}$ all Feynman diagrams obtained by replacing a superficially divergent 1PI graph $\gamma$ by $U({\gamma})$, which is a polynomial in masses and momenta. 
 
 In order to see the connection to the rational $R_2$ terms at one- and two-loops from refs.~\cite{Ossola:2008xq,Pozzorini:2020hkx}, we first split the character $U$ into a sum of two contributions
 \beq\label{eq:UZR}
 U = Z + {\tR}\,,
 \eeq
 where $Z$ is the character that defines the UV counterterms from eq.~\eqref{eq:Birkhoff_renorm}. Injecting eq.~\eqref{eq:UZR} into eq.~\eqref{eq:U_def}, we find the following recursive expression for $\tR$:
 \beq\label{eq:Rtilde}
 \tR = \bF\Big[\Phibar_R - \Phi_R - m_A(\tR\otimes \Phi)\Delta'\Big]\,.
 \eeq
 Note that, since $U(\Gamma)$ and $Z(\Gamma)$ are non-zero only for superficially-divergent graphs and evaluate to polynomials in the momenta and masses, the same is true for $\tR$. 
 We now argue that $ \tR(\Gamma)$ agrees with the definition of the vertices $\delta R_{\Gamma}$ for rational $R_2$ terms of UV origin defined in eqs.~\cite{Ossola:2008xq,Pozzorini:2020hkx} (see also eq.~\eqref{eq:two-loop-R2-master}).  At one-loop, we obtain:
\beq\label{eq:Rtilde_def}
\tR({\Gamma}) = \bF\Big[\Phibar_R(\Gamma) -\Phi_R({\Gamma})\Big] = \bR\cAbar_{\Gamma} - \bR\cA_{\Gamma} +\ord(\eps)= \delta R_{\Gamma}+\ord(\eps) = \delta R_{\Gamma}\,,
\eeq
where the second equality follows from the fact that in the HV scheme, the one-loop UV counterterms for $\cAbar_{\Gamma}$ and $\cA_{\Gamma}$ are identical, and the last equality follows because neither $R_2(\Gamma)$ nor $\delta R_{\Gamma}$ have higher-order terms in $\eps$. At two-loop order, eq.~\eqref{eq:master1} reduces to
\beq\bsp
\bR\cAbar_{\Gamma} 
&\,= \cA_{\Gamma} +\bU\cAbar_{\Gamma} +\sum_{\gamma\in W(\gamma)}\bU\cAbar_{\gamma} \ast \cA_{\Gamma/\gamma}+\ord(\eps)\\
&\,= \cA_{\Gamma} +Z({\Gamma}) + \tR({\Gamma}) +\sum_{\gamma\in W(\gamma)}(Z({\gamma})+\tR({\gamma}))\ast \cA_{\Gamma/\gamma}+\ord(\eps)\\
&\,= \cA_{\Gamma} +\delta Z_{\Gamma} + \tR({\Gamma}) +\sum_{\gamma\in W(\gamma)}(\delta Z_{\gamma}+\tR(\gamma))\ast \cA_{\Gamma/\gamma}+\ord(\eps)\,.
\esp\eeq
Comparing to eq.~\eqref{eq:two-loop-R2-master}, we see that we have $\delta R_\Gamma = \tR({\Gamma})$ also at two loops.
We thus see that the map $R_2$ agrees with the definition of vertices for rational $R_2$ terms of UV origin through two loops form the literature. Explicit examples of what these vertices look like can be found in refs.~\cite{Ossola:2008xq,Draggiotis:2009yb,Garzelli:2009is,Pozzorini:2020hkx,Lang:2020nnl,Lang:2021hnw}. The equations~\eqref{eq:U_def},~\eqref{eq:Phibar0} and~\eqref{eq:Rtilde}, however, are general and allow us to define rational $R_2$ terms of UV origin to arbitrary loop order.
The corresponding Feynman diagrams can be grouped into three classes:
\begin{itemize}
\item Diagrams with only UV counterterm vertices $Z({\gamma})$ inserted: those correspond precisely to the UV counterterms introduced by the BPHZ formula.
\item Diagrams with only $R_2$ vertices $\tR({\gamma})$ inserted.
\item  Diagrams with both UV counterterms and $R_2$ vertices inserted. Those diagrams appear for the first time for 1PR diagrams at two loops, and 1PI diagrams at three loops.
\end{itemize}

We conclude this section by discussing some general properties of the rational $R_2$ terms at arbitrary loop order. First, we can extract the poles of $\tR(\Gamma)$ by acting with $\bK$. Since $\bK\Phibar_R=0$ and $\bK\bF=\bK$, we find
\beq
\bK\tR= -\bK\Big[m_A(\tR\otimes \Phi)\Delta'\Big]\,.
\eeq
We see that the poles of $\tR({\Gamma})$ are entirely determined by lower-loop information. In particular, if $\Gamma$ is a one-loop graph, $\bK\tR({\Gamma}) = 0$. Starting from two loops, $\tR({\Gamma})$ will in general contain poles in $\eps$, in agreement with the results from ref.~\cite{Pozzorini:2020hkx,Lang:2020nnl,Lang:2021hnw}.
Second, unlike $U$ and $Z$, $\tR$ is not a character. This effect did not arise in the two-loop discussion of ref.~\cite{Pozzorini:2020hkx}, because for two-loop 1PI graphs, there are no subdiagrams that are disconnected 1PI graphs. The fact that $\tR$ is not a character comes into play for the first time at three loops for 1PI graphs, and at two loops for 1PR graphs. For example, if $\Gamma=\Gamma_1\Gamma_2$ is a disconnected two-loop graph which is a product of two one-loop graphs, we have
\beq\bsp
\tR({\Gamma_1}\,{\Gamma_2})&-\tR({\Gamma_1})\,\tR({\Gamma_2}) \\
&= U({\Gamma_1}\,{\Gamma_2})-Z({\Gamma_1}\,{\Gamma_2})-\Big(U({\Gamma_1})-Z({\Gamma_1})\Big)\Big(U({\Gamma_2})-Z({\Gamma_2})\Big)\\
&=U(\Gamma_1)\,Z(\Gamma_2)-Z(\Gamma_1)\,U(\Gamma_2)\,,
\esp\eeq
and so $\tR({\Gamma_1}\,{\Gamma_2})\neq\tR({\Gamma_1})\,\tR({\Gamma_2})$ in general. Hence, $\tR$ is not a character.

Finally, we need to comment on IR divergencies. Strictly speaking, our derivation only holds in the absence of IR divergencies. It is only proven that rational $R_2$ terms of IR origin cancel at one loop in the HV scheme
at the level of the full amplitude~\cite{Bredenstein:2008zb}. In a theory with IR divergencies, in particular massless gauge theories, we may still use the $\tR$ map to extract the contribution to the rational $R_2$ terms that are of UV origin. In order to isolate those UV contributions, we may, e.g., proceed as in ref.~\cite{Pozzorini:2020hkx} and apply a tadpole decomposition~\cite{Misiak:1994zw,Chetyrkin:1997fm} to the integrand. This results in a sum of integrals with massive propagators, which are free of IR divergencies, and a contribution with negative superficial degree of divergence, which does not contribute to the $R_2$ terms of UV origin. In this way we can isolate the UV divergent contributions to an integral, and extract the corresponding rational $R_2$ terms using the $\tR$ map.



\section{Conclusion}
\label{sec:conclusion}

In this paper we have shown that is possible to define a set of process-independent vertices, that are polynomial in momenta and masses, such that rational $R_2$ terms of UV origin can be obtained by computing all possible lower-loop diagrams where some superficially divergent subgraphs have been replaced by these vertices. Our result is remarkably simple, and shows that vertices for rational $R_2$ terms satisfy, to all orders in perturbation theory, exactly the same properties as UV counterterms in the $\MS$-scheme. This gives a field theoretic interpretation and definition of these vertices for rational $R_2$ terms to all orders.

So far our result is restricted to rational $R_2$ of UV origin. At one loop it is known that all rational terms of infrared (IR) origin cancel in the final result for the scattering amplitude in the HV scheme. At two loops and beyond, it is still an open question whether or not these rational terms of IR origin cancel. For the future, it would be interesting to investigate if the cancellation observed at one loop persists at higher loops, and if not, if a similar all-order field-theoretic definition as in the UV case can be obtained.


\section*{Acknowledgments}
This project is supported by the DFG project 499573813 ``EFTools''.

\appendix

\section{Proof of the main theorem}
\label{sec:proof}

In this appendix we present proof of Theorem~\ref{thm:main}. The proofs of Properties 2, 3 \& 4 are 
 similar to the corresponding proofs for $\bRbar$ given in ref.~\cite{Caswell:1981ek} and proceed by induction in the number of loops of $\Gamma$, and all properties are easily seen to hold for one-loop graphs. Since Properties 2, 3 \& 4 also hold for the character $Z$ defining the UV counterterms, it is sufficient to prove Properties 2, 3 \& 4 for the map $\tR$.

\paragraph{Proof of Property 1.} We start by showing that $U$ is a character. The proof proceeds by induction in the number of loops. 
Assume that $\Gamma=\Gamma_1\Gamma_2$. If $\Gamma$ is a one-loop graph, then either $\Gamma_1$ or  $\Gamma_2$ is the trivial graph. We assume without loss of generality that $\Gamma_2$ is the trivial graph. Then we clearly have $U(\Gamma_1\Gamma_2) = U(\Gamma_1)U(\Gamma_2)$. 
In order to proceed, we note that the fact that $\Phibarzero(\Gamma)$ is always finite implies that also $(U\star \Phi)(\Gamma)$ is finite. Moreover, since $\Phibarzero$ is a character, we have
\beq\bsp\label{eq:proof01}
0 &\,= \Phibarzero(\Gamma_1\Gamma_2) - \Phibarzero(\Gamma_1)\Phibarzero(\Gamma_2)\\
&\,=\bF\big[(U\star \Phi)({\Gamma_1}\Gamma_2)\big] -\bF\big[(U\star \Phi)({\Gamma_1})\big]\,\bF\big[(U\star \Phi)({\Gamma_2})\big]\\
&\,=\bF\big[(U\star \Phi)({\Gamma_1}\Gamma_2)-(U\star \Phi)({\Gamma_1})(U\star \Phi)({\Gamma_2})\big]\,,
\esp\eeq
where the last step follows form the fact that $\bF[ab] = \bF[a]\bF[b]$ whenever $a$ and $b$ are finite.

It is easy to see that $\bW(\Gamma) = \bW(\Gamma_1)\times\bW(\Gamma_2)$. This implies
\beq\bsp\label{eq:mult_1}
(U\star \Phi)({\Gamma})&\, =  \sum_{\gamma_1\in \bW(\Gamma_1)}\sum_{\gamma_2\in \bW(\Gamma_2)}U({\gamma_1\gamma_2})\ast\big(\Phi({\Gamma_1/\gamma_1})\,\Phi({\Gamma_2/\gamma_2})\big)\,.
\esp \eeq
We also have
\beq\bsp
(U\star \Phi)({\Gamma_1})\,&(U\star \Phi)({\Gamma_2}) =\\
&\,= \sum_{\gamma_1\in \bW(\Gamma_1)}\sum_{\gamma_2\in \bW(\Gamma_2)}
\big(U({\gamma_1})\ast\Phi({\Gamma_1/\gamma_1})\big)\,\big(U({\gamma_2})\ast\Phi({\Gamma_2/\gamma_2})\big)\,.
\esp\eeq
We can have a closer look at the terms in the last sum. If we denote the number of loops of $\Gamma_i/\gamma_i$ by $L_i$ ($i=1,2$), we have
\beq\bsp
\big(U({\gamma_1})&\ast\Phi({\Gamma_1/\gamma_1})\big)\,\big(U({\gamma_2})\ast\Phi({\Gamma_2/\gamma_2})\big) =\\
&\,=
\int\left(\prod_{i=1}^{L_1}\frac{\rd^D\bar{q}_i}{(2\pi)^D}\right)\,\left(\prod_{i=1}^{L_2}\frac{\rd^D\bar{q}'_i}{(2\pi)^D}\right)\,U({\gamma_1})\,U({\gamma_2})\,\frac{\cN_{\Gamma_1/\gamma_1}\,\cN_{\Gamma_2/\gamma_2}}{\prod_e\overline{D}_e}\\
&\,=
\int\left(\prod_{i=1}^{L_1+L_2}\frac{\rd^D\bar{q}_i}{(2\pi)^D}\right)\,U({\gamma_1})\,U({\gamma_2})\,\frac{\cN_{\Gamma/(\gamma_1\gamma_2)}}{\prod_e\overline{D}_e}\\
&\,=\big(U({\gamma_1})\,U({\gamma_2})\big)\ast\Phi({\Gamma/(\gamma_1\gamma_2)})\,,
\esp\eeq
where we used the fact that $\cN_{\Gamma_1/\gamma_1} \cN_{\Gamma_2/\gamma_2} = \cN_{\Gamma/(\gamma_1\gamma_2)}$.
Hence,
\beq\bsp\label{eq:mult_2}
(U\star \Phi)({\Gamma_1})\,(U\star \Phi)({\Gamma_2}) &\, =\sum_{\gamma_1\in \bW(\Gamma_1)}\sum_{\gamma_2\in \bW(\Gamma_2)}\big(U({\gamma_1})\,U({\gamma_2})\big)\ast\Phi({\Gamma/(\gamma_1\gamma_2)})\,.
\esp\eeq
Inserting eqs.~\eqref{eq:mult_1} and~\eqref{eq:mult_2} into eq.~\eqref{eq:proof01}, we see that all terms cancel, except the one for $(\gamma_1,\gamma_2)=(\Gamma_1,\Gamma_2)$, and we have
\beq
\bF\big[U(\Gamma_1\Gamma_2) - U(\Gamma_1)U(\Gamma_2)\big]=0\,.
\eeq
Since $\bF$ is a projector, we have $\bF\big[U(\Gamma_1\Gamma_2)\big] = U(\Gamma_1\Gamma_2)$. Moreover, the Rota-Baxter property of $\bF$ implies that
\beq
\bF[a]\bF[b] = \bF[\bF[a]b]+\bF[a\bF[b]] - \bF[ab]\,,
\eeq
and so, using the fact that $\bF$ is a projector,
\beq
\bF\big[\bF[a]\bF[b]\big] = \bF[\bF[a]b]+\bF[a\bF[b]] - \bF[ab]=\bF[a]\bF[b]\,.
\eeq
From this it follows that $\bF\big[U(\Gamma_1)U(\Gamma_2)\big]=U(\Gamma_1)U(\Gamma_2)$. Hence $U(\Gamma_1\Gamma_2) - U(\Gamma_1)U(\Gamma_2)=0$, and so $U$ is a character.

\paragraph{Proof of Property 2.} We now show that if $\omega(\Gamma)<0$, then $\tR({\Gamma})=0$. The proof proceeds by induction in the number of loops of $\Gamma$. 

If $\omega(\Gamma)<0$, then it may still contain superficially divergent subgraphs. It is easy to show that in that case $\Gamma$ always has a set of 1PI subgraphs $\gamma_1,\ldots,\gamma_k$ such that all divergent subgraphs are contained in their product $\gamma=\gamma_1\cdots\gamma_k$~\cite{Caswell:1981ek}. The map $b$ from eq.~\eqref{eq:b_def} has the property~\cite{Caswell:1981ek}:
\beq
b(\Gamma)= \Phibar_R(\gamma)\ast\Phibar(\Gamma/\gamma)\,,\qquad \omega(\Gamma)<0\,.
\eeq
By Weinberg's theorem, the integral on the right-hand side is convergent, so that 
\beq
\Phibar_R({\Gamma}) = \Phibar_R(\gamma)\ast\Phibar(\Gamma/\gamma)\,,\qquad \omega(\Gamma)<0\,.
\eeq
Exactly the same relation holds if we replace $\Phibar$ by $\Phi$ everywhere. This gives
\beq\bsp\label{eq:proof1}
\tR(\Gamma)&\,= \bF\big[\Phibar_R(\gamma)\ast\Phibar(\Gamma/\gamma) -\Phi_R(\gamma)\ast\Phi(\Gamma/\gamma) - m_A(\tR\otimes \Phi)\Delta'(\Gamma)\big]\\
&\,= \bF\big[\Phibar_R({\gamma})\ast \widetilde{\Phi}({\Gamma/\gamma}) \big]  + \bF\big[X({\gamma}) \ast \Phi({\Gamma/\gamma})\big]
\,,
\esp\eeq
where we defined
\beq
\widetilde{\Phi}({\Gamma/\gamma})= \Phibar({\Gamma/\gamma}) - {\Phi}({\Gamma/\gamma})\,,
\eeq
and
\beq
X({\gamma}) =\Phibar_R({\gamma}) - \Phi_R({\gamma}) -\tR({\gamma})- m_A(\tR\otimes\Phi)\Delta'(\gamma)\,,
\eeq
 and we used the induction hypothesis, which implies that $\tR({\gamma'})=0$, unless $\gamma'\subset \gamma$ (because all divergent subgraphs must lie in $\gamma$). By definition of $\tR(\gamma)$, we have $X({\gamma})=\ord(\eps)$. The integral
\beq
X({\gamma})\ast {\Phi}({\Gamma/\gamma}) = \int\left(\prod_{i=1}^{L}\frac{\rd^D\bar{q}_i}{(2\pi)^D}\right)\,X({\gamma})\,\frac{{\cN}_{\Gamma/\gamma}}{\prod_{e}\overline{D}_e}
\eeq
is convergent (because $\Gamma/\gamma$ does not contain any subdivergences), and so $X({\gamma})\ast {\Phi}({\Gamma/\gamma})  = \ord(\eps)$, or equivalently $\bF\big[X({\gamma})\ast {\Phi}({\Gamma/\gamma})\big]=0$.
Similarly, we have, with $\widetilde{\cN}_{\Gamma/\gamma}= \overline{\cN}_{\Gamma/\gamma} - {\cN}_{\Gamma/\gamma}$,
\beq
\Phibar_R({\gamma})\ast \widetilde{\Phi}({\Gamma/\gamma}) = \int\left(\prod_{i=1}^{L}\frac{\rd^D\bar{q}_i}{(2\pi)^D}\right)\,\Phibar_R({\gamma})\,\frac{\widetilde{\cN}_{\Gamma/\gamma}}{\prod_{e}\overline{D}_e}\,.
\eeq
The integral is convergent, and so it can be evaluated in four dimensions, where $\widetilde{\cN}_{\Gamma}$ vanishes. Hence, $\Phibar_R({\gamma})\ast \widetilde{\Phi}({\Gamma/\gamma}) = \ord(\eps)$, and so $\bF\big[\Phibar_R({\gamma})\ast \widetilde{\Phi}({\Gamma/\gamma})\big]=0$. We conclude that both terms in eq.~\eqref{eq:proof1} vanish, and so $\tR({\Gamma})=0$.

\paragraph{Proof of Property 3.} 
The proof follows exactly the same lines as for the $\Phi_R$ map in ref.~\cite{Caswell:1981ek}. In essence, the definition of the $\tR$ map is purely combinatorial. Derivatives instead act on the analytic expressions, and commute with the combinatorial formulas. We detail the proof here for completeness.


Let us compute $\partial_a\tR({\Gamma})$. We proceed by induction in the number of loops. Because of eqs.~\eqref{eq:Phi_der} and~\eqref{eq:Phi_R_der}, we focus on the last term in eq.~\eqref{eq:Rtilde}, for which we introduce the notation
\beq
T := m_A(\tR\otimes \Phi)\Delta'\,.
\eeq
We thus need to show that $T$:
\beq
\partial_aT = T\delta_a\,.
\eeq
We have:
\beq\bsp
\partial_aT&\,
=\partial_am_A(\tR\otimes \Phi)\Delta'\\
&\,=m_A(\partial_a\otimes\id+\id\otimes \partial_a)(\tR\otimes \Phi)\Delta'\\
&\,=m_A(\partial_a\tR\otimes\Phi +\tR\otimes \partial_a\Phi)\Delta'\\
&\,=m_A(\tR\otimes \Phi)(\delta_a\otimes\id+\id\otimes \delta_a)\Delta'\\
&\,=m_A(\tR\otimes \Phi)\Delta'\delta_a\\
&\,=T\delta_a\,.
\esp\eeq
where the fourth step follows from the induction hypothesis. 

\paragraph{Proof of Property 4.} The proof of Property 4 goes along the same lines as the corresponding proof for the UV counterterms in ref.~\cite{Caswell:1981ek}.

It is easy to see that, if $P$ is any function in external momenta and internal masses, then $P$ is a polynomial of degree at most $d$ if and only if for all $s>d$ and for all $\underline{a}=(a_1,\ldots,a_s)$, $P$ is annihilated by the differential operator $\cD_{\underline{a}} = \partial_{a_1}\cdots \partial_{a_s}$. In the following we will use the notation $s =|\underline{a}|$.

The numerator $\cNbar_{\Gamma}$ is by definition a polynomial in external momenta and internal masses. Let $d_{\cNbar}$ be the degree of this polynomial. It follows that for all $s>d_{\cNbar}$ and $\underline{a}=(a_1,\ldots,a_s)$, we have 
\beq\label{eq:proof3_1}
\cD_{\underline{a}}\cNbar_{\Gamma} = 0\,.
\eeq
Let us now act with such an operator on $\cAbar_{\Gamma}$. If a derivative $\partial_{a_i}$ acts on a propagator, then the result is either zero, or we obtain an expression with strictly  lower superficial degree of divergence. Indeed, in the case where the result of the differentiation is not zero, we have (cf.~eq.~\eqref{eq:pro_Def}):
\beq\bsp
\frac{\partial}{\partial p_i^\mu}\frac{1}{\overline{D}_e} &\,= -\frac{2\alpha_{e,i}}{\overline{D}_e^2}\,\left(\sum_{i=1}^n\alpha_{e,i}\,p_{i\mu}+\sum_{j=1}^L\beta_{e,j}\,\overline{q}_{j\mu}\right)\,,\\
\frac{\partial}{\partial m_e}\frac{1}{\overline{D}_{e'}} &\,= -\frac{2}{\overline{D}_{e'}^2}\,\frac{\partial m_{e'}}{\partial m_e}\,.
\esp\eeq
It is easy to see that the expressions on the right-hand side have an improved behaviour at large loop momentum. 
As a consequence, if we act with $\cD_{\underline{a}}$ and use the Leibniz rule, all terms have a strictly lower superficial degree of divergence, except for the case where no derivative acts on any propagator. That term corresponds to the situation where $\cD_{\underline{a}}$ acts on $\cNbar_{\Gamma}$, in which case we get zero by eq.~\eqref{eq:proof3_1}. Hence, we conclude for all $\underline{a}$ with $|\underline{a}|>d_{\overline{\cN}}$, we have $
\omega(\delta_{\underline{a}}\Gamma)  < \omega(\Gamma)$, with $\delta_{\underline{a}} = \delta_{a_1}\cdots\delta_{a_s}$.
We can continue this way, and we obtain a sequence of positive integers $d_{\cNbar}=d_1, d_2,\ldots, d_k$ such that for all $\underline{a}$ with $|\underline{a}|>d_{\Gamma} = \sum_{i=1}^kd_i$, we have $\omega(\delta_{\underline{a}}\Gamma)  < 0$.
In other words, we have shown that for every graph $\Gamma$ there is a positive integer $d_{\Gamma}$ such that for all $\underline{a}$ with $|\underline{a}|>d_{\Gamma}$, $\omega(\delta_{\underline{a}}\Gamma)<0$. 

Next, consider a graph $\Gamma$ with $\omega(\Gamma)>0$. From Properties 2 \& 3, we have, for all $\underline{a}$ with $|\underline{a}|>d_{\Gamma}$,
\beq
\cD_{\underline{a}}\tR({\Gamma}) = 
\tR\big(\delta_{\underline{a}}\Gamma\big)  = 0\,,
\eeq
 and so $\tR({\Gamma})$ is a polynomial of degree at most $d_{\Gamma}$ in external momenta and propagator masses.

\bibliographystyle{JHEP}
\bibliography{bib}

\end{document}